\documentclass[%
 reprint,
 amsmath,amssymb,
 aps,
prb,
]{revtex4-2}
\usepackage{filecontents}
\usepackage{graphicx}
\usepackage{dcolumn}
\usepackage{bm}
\newcommand{\eu}{EuB$_6$}
\newcommand{\sm}{SmB$_6$}
\newcommand{\sro}{Sr$_3$Ru$_2$O$_7$}
\newcommand{\andrea}{\textcolor{black}}
\usepackage{times}
\usepackage{amsmath}
\usepackage{fouriernc}
\usepackage{latexsym}
\usepackage{physics}
\usepackage{xcolor}

\begin{document}

\preprint{APS/123-QED}

\title{\andrea{Possible} quantum nematic in a colossal magnetoresistance material}
\author{G. Beaudin}
\affiliation{D\'epartement de Physique, Universit\'e de Montr\'eal, Montr\'eal, Canada}
\altaffiliation{Regroupement Qu\'eb\'ecois sur les Mat\'eriaux de Pointe (RQMP)}
\author{L. M. Fournier}
\affiliation{D\'epartement de Physique, Universit\'e de Montr\'eal, Montr\'eal, Canada}
\altaffiliation{Regroupement Qu\'eb\'ecois sur les Mat\'eriaux de Pointe (RQMP)}
%
%
\author{M. Nicklas}
\affiliation{Max Planck Institute for Chemical Physics of Solids, Dresden, Germany}
\author{M. Kenzelmann}
\affiliation{Laboratory for Neutron Scattering and Imaging, Paul Scherrer Institut, Villigen, Switzerland}
\author{M. Laver}
\affiliation{Department of Physics, University of Warwick, Coventry, UK}
\author{W. Witczak-Krempa}
\affiliation{D\'epartement de Physique, Universit\'e de Montr\'eal, Montr\'eal, Canada}
\altaffiliation{Regroupement Qu\'eb\'ecois sur les Mat\'eriaux de Pointe (RQMP)}
\affiliation{
Centre de Recherches Math\'ematiques, Universit\'e de Montr\'eal, Montr\'eal, Qu\'ebec, Canada}

\author{A. D. Bianchi}
\affiliation{D\'epartement de Physique, Universit\'e de Montr\'eal, Montr\'eal, Canada}
\altaffiliation{Regroupement Qu\'eb\'ecois sur les Mat\'eriaux de Pointe (RQMP)}

\
\date{\today}

\begin{abstract}

\andrea{\eu\ has for a long time captured  the attention of the physics community, as it shows a ferromagnetic phase transition leading to a insulator the metal transition together with  colossal magnetoresistance (CMR). \eu\ has a very low carrier density, which is known to drastically change the interaction between the localised Eu moments and the conduction electrons. One of early triumphs of the quantum theory in condensed matter was the presence of Fermi surface, which is intimately linked to the symmetry of the underlying crystal lattice. This symmetry can be probed by angle resolved magnetoresistance (AMRO) measurements. }
Here, we present angle resolved magnetoresistance (AMRO) measurements that show a that in \eu\ this symmetry is broken, possibly indicating the presence of  a quantum nematic phase. We identify the region in the temperature-magnetic field phase diagram where the magnetoresistance shows two-fold oscillations instead of the expected fourfold pattern.  Quantum nematic phases are analogous to classical liquid crystals. Like liquid crystals, which break the rotational symmetry of space, their quantum analogs break the point-group symmetry of the crystal due to strong electron-electron interactions, as in quantum Hall states, Sr$_3$Ru$_2$O$_7$, and high temperature superconductors.  This is the same region where magnetic polarons were previously observed, suggesting that they drive the nematicity in \eu. This is also the region of the phase diagram where \eu\ shows a colossal magnetoresistance (CMR). This novel interplay between magnetic and electronic properties could thus be harnessed for spintronic applications.\end{abstract}


\maketitle

\section{Introduction}

One of the great successes of the quantum theory of solids was the finding that electrons in crystals largely behave as a quantum gas of free particles. It was soon understood that the interactions between electrons can change their behavior into that of a quantum liquid leading among other things to a modified mass of the quasi-particles. However, unlike in classical liquids, where a suspension of rod-like molecules can lead to anisotropic interactions and the occurrence of nematic phases in liquid crystals, the point like nature of electrons and their interactions seems at first not to lend itself to the formation of a quantum nematic. So, it was quite a surprise when experiments, first in ultraclean quantum Hall systems~\cite{lilly_evidence_1999} and later in \sro\cite{borzi_formation_2007}, indicated the presence of an electronic nematic phase.

Strong electronic correlations are believed to be at the source of theses exotic electronic liquids, as theoretically first predicted for the case of a doped two-dimensional Mott insulator~\cite{kivelson_electronic_1998}  \textcolor{black}{due to the melting of a striped phase}, and later for quantum Hall systems~\cite{macdonald_quantum_2000}. Quantum nematics were also discovered in the high temperature cuprate superconductors~\cite{cheong_incommensurate_1991,tranquada_coexistence_1997,ando_electrical_2002,hinkov_electronic_2008,ramshaw2017highTc}, as well as in the iron arsenide superconductors~\cite{chuang_nematic_2010,Fernandes2014,Watson_2015}, where the relation between the nematic order and superconductivity, and its relation to the close-by structural instability are hotly debated. Pomeranchuk was the first to describe the mechanism by which a Fermi surface can spontaneously break the rotational symmetry~\cite{pomeranchuk_stability_1959}.

This mechanism has been invoked in the case of the cuprate superconductors~\cite{dellanna_electrical_2007,yamase_fermi-surface_2012}, and it may also play a role in the iron arsenides~\cite{chuang_nematic_2010}. Nematic order is also found in CeRhIn$_5$, a heavy fermion superconductor, where it is most likely related to a spin texture~\cite{Ronning2017}. Nematicity found in CeB$_6$~\cite{demishev_ceb6_2017} suggests that a nematic state can be observed in 3D materials such as the hexaborides, and is not only linked to 2D and quasi-2D systems, opening possibilities for more complex quantum materials. Here, we describe a novel type of nematic order in the hexaboride \eu, which due to CMR effects associated with the nematic phase carries the potential to play a new role in spintronics. In spintronics, the spin degrees of freedom are used to obtain transistor action, leading to the promise of a lower energy consumption and the unification of storage and processing components. 

The interplay between the electronic and magnetic properties of \eu\ are still the subject of controversy despite its simple cubic crystal structure $(Pm\overline{3}m)$. \eu\ has a very low charge carrier density~\cite{aronson_fermi_1999, zhang_spin-dependent_2008} ($\approx 10^{19}~\mbox{cm}^{-3}$), these carriers couple to localized Eu $4f$ moments which are pure spin with $S=7/2$, but whether it should be considered as a semimetal or semiconductor is still an ongoing debate~\cite{massidda_electronic_1996,tromp_cab6_2001,kim_spin-split_2013}. \eu\ becomes ferromagnetic at 
$T_\mathrm{C} = 12.6$~K, accompanied by an order of magnitude reduction in resistivity and CMR~\cite{sullow_structure_1998} in the vicinity of $T_\mathrm{C}$. The specific heat of \eu\ shows an additional anomaly at $T_M = 15.5$~K~\cite{degiorgi_low-temperature_1997, sullow_structure_1998,sullow_metallization_2000}. This temperature coincides with the first anomaly in the electrical resistivity. In the literature, this anomaly  in the electrical resistivity is typically associated with the percolation transition of  magnetic polarons. At this temperature,  the polarons begin to overlap, releasing the trapped charge carriers which consequently lowers the resistivity~\cite{pohlit_evidence_2018}.

Magnetic polarons are expected to be important in \eu\ due to its low carrier density and they were indirectly indicated by a number of experiments~\cite{p.nyhus_spectroscopic_1997,sullow_metallization_2000,brooks_magnetic_2004,zhang_nonlinear_2009,das_magnetically_2012,manna_lattice_2014, kim_optical_2008}. Heuristically, magnetic polarons are composite objects that form when charge carriers polarize a puddle of local moments and become trapped in that puddle~\cite{wegener_fluctuation-induced_2002}. This mechanism is thought to be behind the large CMR effect in \eu. Since the merger of these polarons leads to a substantially enhanced charge mobility, magnetic polarons in \eu\ were directly identified through a small angle neutron scattering (SANS) experiment~\cite{beaudin_2019}, while scanning tunneling microscopy (STM) shows that \eu\ is electronically inhomogeneous in the same temperature region~\cite{pohlit_evidence_2018}. True ferromagnetic order, however, is established only at $T_\mathrm{C}$~\cite{sullow_structure_1998,henggeler_magnetic_1998}. This scenario with a transition in two steps is supported by various experimental techniques, such as resistivity and magnetization measurements~\cite{sullow_structure_1998,wigger_electronic_2004}, nonlinear Hall effect~\cite{zhang_nonlinear_2009}, or muon-spin rotation~\cite{brooks_magnetic_2004}. These previous experiments all show a linear upward trend of the phase boundary for the phase diagram due to magnetic polarons. In particular, the SANS experiment provided clear evidence for the presence of magnetic polarons in \eu~\cite{beaudin_2019}.

In this paper, we use angle-dependent magnetoresistance oscillations (AMRO) to map out the quantum nematic phase in \eu\ and show that it exists in the region of the phase diagram where magnetic polarons are observed, \andrea{just below the phase line $T_\mathrm{QN}$ in the $(H, T)$ phase diagram, as indicated by magnetostriction measurements~\cite{manna_lattice_2014} and by our measurements of the complex part $\chi''$ of the magnetic AC susceptibility $ \chi_\mathrm{AC} = \chi' - i\chi"$.}

The AMRO technique has been previously used with particular success in the case of metals with two-dimensional Fermi surfaces, such as organic conductors~\cite{yamaji1989,Yagi1990} and Sr$_2$RuO$_4$~\cite{Bergemann2003}. Later, this technique was successfully used to demonstrate the presence of a Fermi surface in the overdoped high temperature superconductor Tl$_2$Ba$_2$CuO$_{6+\delta}$~\cite{Hussey_highTc_2003}, as well as in YBa$_2$Cu$_3$O$_{6.58}$~\cite{ramshaw2017highTc}. As the Fermi surface of most materials is not spherical, including the one of \eu, the electrons no longer move on circular orbits. This leads to an angular dependence of the magnetoresistance and consequently to angle-dependent oscillations in the magnetoresistance. In \eu\, previous studies showed a four-fold AMRO pattern at high magnetic fields~\cite{urbano_magnetic_2004}. A two-fold AMRO pattern breaking the axis symmetry was subsequently reported by Glushkov \emph{et al.}~\cite{Glushkov_2009}, however this was attributed to demagnetization effects. Here, measuring multiple samples allowed us to separate the demagnetization  effects from the  AMRO signal, and to definitively determine the presence of an additional AMRO component, which suggests the presence of a quantum nematic in \eu.

\section{Methods}
The \eu\ single crystals used in our study were grown by the same method as previously used for the sample in Refs.~\cite{degiorgi_low-temperature_1997,sullow_structure_1998,sullow_metallization_2000}.
Single crystals were flux-grown with a ratio of 14~mg of \eu\ powder per gram of Al flux. The mixture was heated in an Al$_2$O$_3$ crucible using a vertical tube furnace to 1500$^{\circ} $C at a rate of 200$ ^{\circ} $C/hour in a flow of high purity Ar. The mixture was held at that temperature for 10~hours and then cooled down to 1000$ ^{\circ} $C at a rate of 5$ ^{\circ} $C/hour \cite{fisk_magnetic_1979}. The crystals were separated from the Al flux in boiling sodium hydroxide.

Specific heat $C_p$ was measured from 0.4 to 30~K using a Quantum Design Physical Properties Measurement System (PPMS) with a $^3$He insert. Magnetic susceptibility was measured using a Quantum Design PPMS AC susceptibility option. The AC susceptibility $\chi_\mathrm{AC}$ is defined by $\chi_\mathrm{AC} = \chi' - i\chi"$. For these measurements, the field was applied along $[111]$ with an oscillating component of 5~Oe at a frequency of 77~Hz.

Resistivity was measured for temperatures between 1.8 and 300~K and magnetic fields between 0 and 9~T with a PPMS rotator using four-point contact with spot-welded gold wires. Measurements of angle-dependent magnetoresistance oscillations (AMRO) were made by rotating the applied magnetic field $H$ in a plane perpendicular to the current. Measurements were made up to $\mu_0 H = 2$~T. At the measured temperatures, no hysteresis was observed. \andrea{The experiment was done on multiple samples with different cross-sections, and aspect ratios. A list of the samples and their dimensions is given in Table~\ref{tab:demag}. An image of the main sample \#1 is shown in the Appendix (Fig.~\ref{fig:sample}).} 

For a sample with a rectangular cross-section perpendicular to the current direction, rotation of the applied magnetic field $H$ in the plane perpendicular to the current leads to a change in the induction $B$ due to demagnetization effects. This effect appears even though the magnitude of $H$ is constant and gives rise to a two-fold AMRO contribution $a_\mathrm{dem}$.

\begin{figure*}
\includegraphics[width=0.7\textwidth]{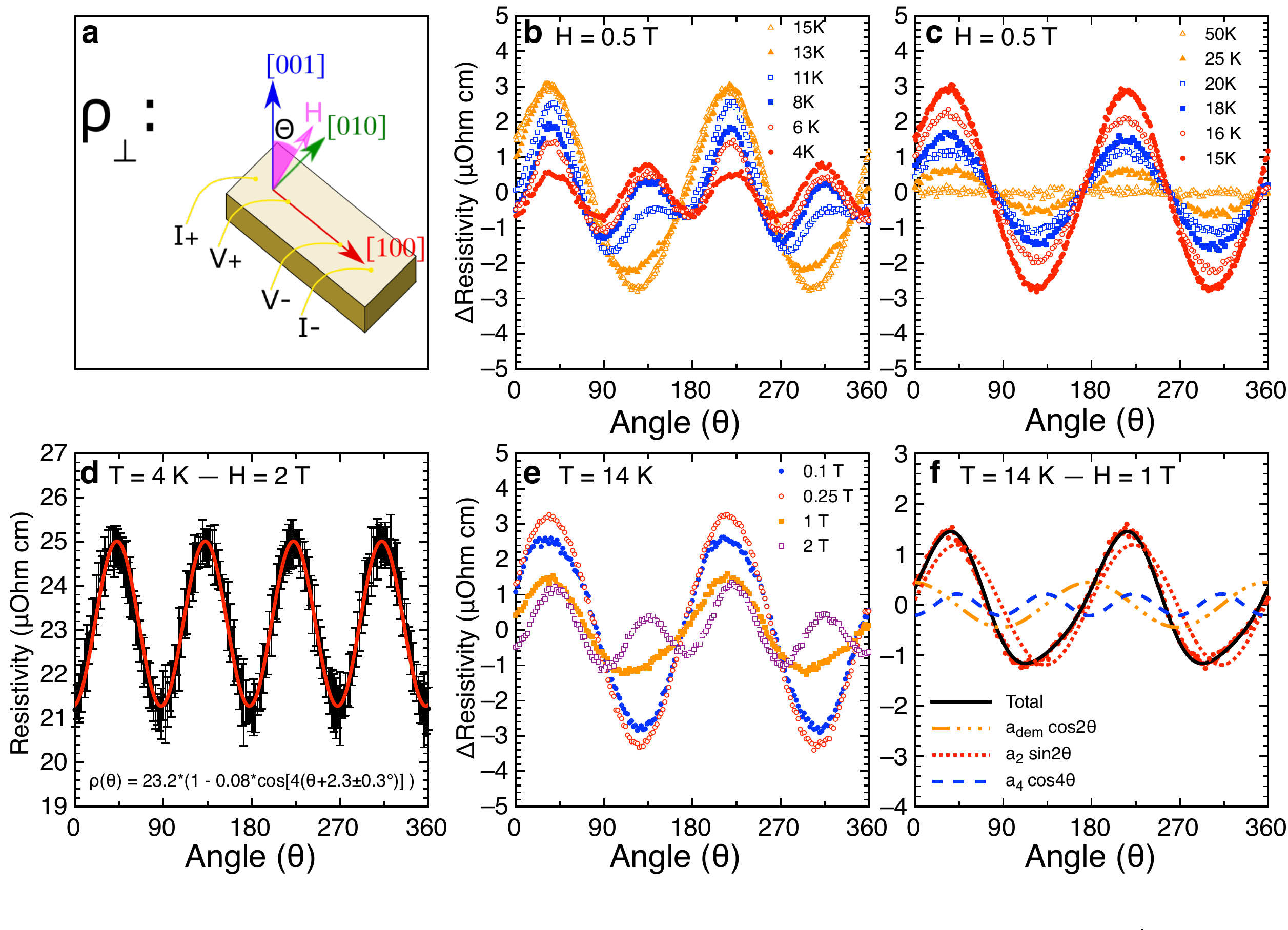}
\caption{\andrea{\textbf{Angle-dependent magnetoresistance oscillations (AMRO) in a sample of almost square cross-section (sample \#1):} a) Schematics of the measurement geometry: for measuring the perpendicular magnetoresistance $\rho_\perp$, the applied magnetic field $H$ is rotated by an angle $\theta$ in the $(100)$-plane from $[001]$ to $[010]$, thus maintaining $H$ perpendicular to the current flowing along $[100]$.  b) and c) show the oscillatory component of $\rho_\perp-\rho_0$ versus the field angle $\theta$ for $\mu_0 \, H=0.5$~T at different temperatures. d) shows the resistivity at 4~K and 2~T used to determine the experimental rotator angle $\theta_0 = -2.3^\circ$ at which $H \parallel [001]$. The solid line is the fit to a fourfold contribution. This figure also shows the typical noise in our AMRO measurements; these uncertainties were taken into consideration when fitting the AMRO oscillations. e) Perpendicular magnetoresistance $\rho_\perp-\rho_0$ versus $\theta$ at $T=14$~K for different applied fields.  f) Comparison of the different AMRO components at 14~K and 1~T, as described by Eq.~\ref{eq:rho_H_T_theta}.
{\it ML Comments on figure panels: Panel a: use $\theta$, not $\Theta$ to label rotation angle. Panels b, c, d and f: add $\mu_0$ before $H$ in legends. Panels d and f: use ``;'' instead of --- between temperature and field in legends.}
}}
    \label{fig:RvsTheta_cubic}
\end{figure*}

\andrea{\begin{table}
    \centering
    \begin{tabular}{c c c c c c c c c}
     Sample & $L$ & $t$ & $w$ & $\mathcal{D}_{010}$/ $\mathcal{D}_{001}$& $B_{001}-B_{010}$ \\  
     \# & (mm) & (mm) & (mm) &  & (mT) \\ 
     \hline
     \hline
    1 & 2.413  & 0.428 & 0.437 & 0.983 & -6.3  \\ 
     2 & 2.040  & 0.590 & 0.562 & 1.046 & 15.5  \\ 
     3 & 7.89 & 0.514 & 0.550 & 0.94 & -23.8  \\ 
    4 & 4.32 & 0.650 & 0.507 & 1.25 & 85.9 \\ 
    5 & 5.090 & 0.459 & 0.648 & 0.73 & -119.2 \\ 
    6 & 1.47 & 0.143 & 0.866 & 0.19 & -501.9 \\ 
     \hline\hline
\end{tabular}
    \caption{\andrea{\textbf{The demagnetization factors for the measured samples.} The second, third, and fourth columns list the sample dimensions, where $L$ is the length along $[100]$,   $t$ the thickness along $[001]$,  and $w$  the width along $[010]$.  The fourth column is the ratio between the demagnetization factor along the two different directions $\mathcal{D}_{010}$/$\mathcal{D}_{001}$. The sixth column is the difference in internal field between the $[001]$ and $[010]$ directions for $H=0.5$~T and $T=15$~K.} }
    \label{tab:demag}
\end{table}}

\section{Results and Discussion}

 \begin{figure}
\includegraphics[width=0.9\linewidth]{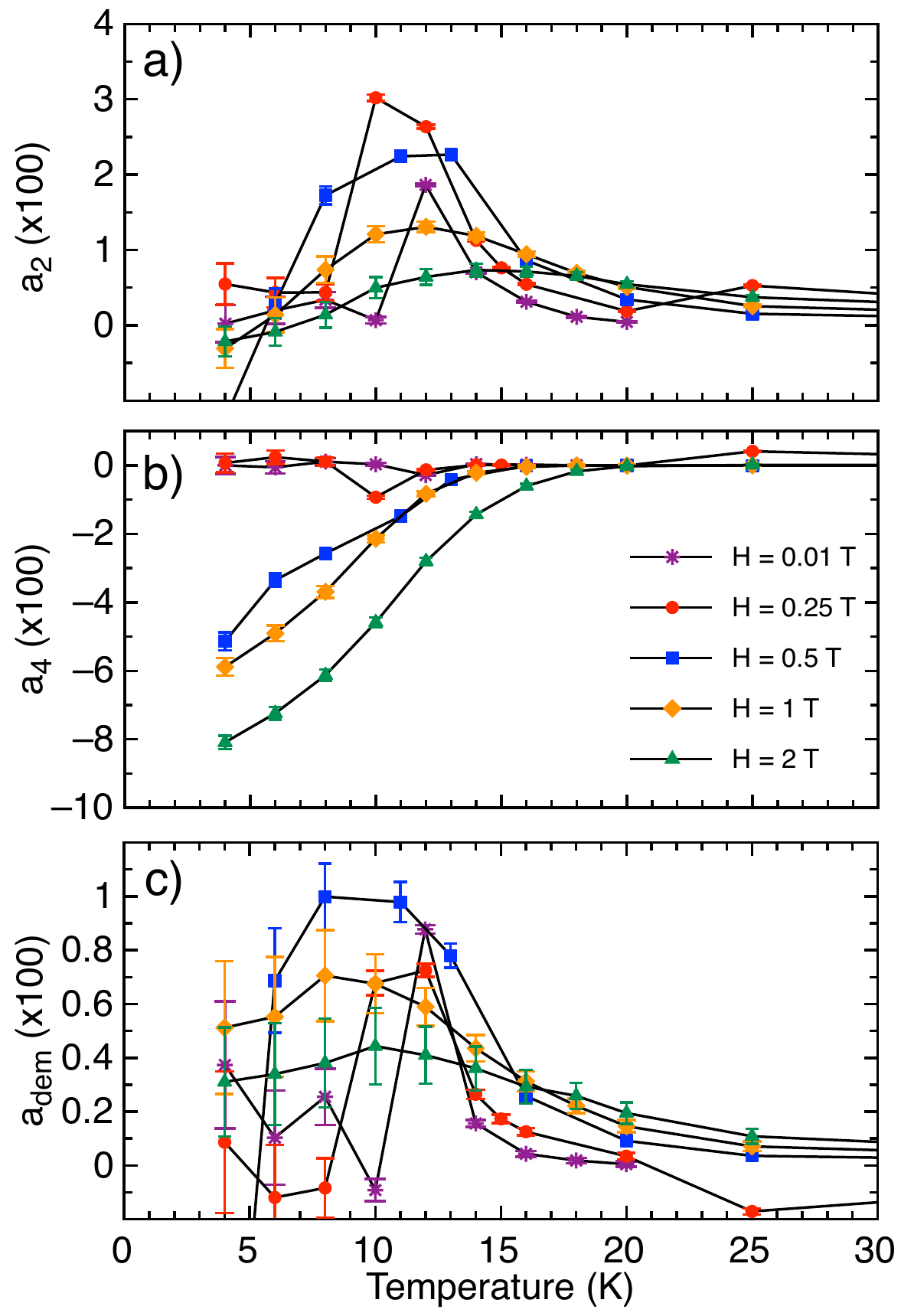}
\caption{\andrea{\textbf{Amplitude of the different AMRO components versus the temperature} a) Amplitude $a_2$ of the two-fold symmetry breaking components, b) Amplitude $a_4$ of the fourfold component due to the shape of the Fermi surface, and c) amplitude $a_{\mathrm{dem}}$ of the demagnetization for applied magnetic fields of 0.01~T (purple stars), 0.25~T (red circles), 0.5~T (blue square), 1~T (orange diamonds) and 2~T (green triangles). This data was taken on sample \#1 in Table~\ref{tab:demag}.} }
\label{fig:amplitudes_vs_T}
\end{figure}


\begin{figure*}
\includegraphics[width=0.8\textwidth]{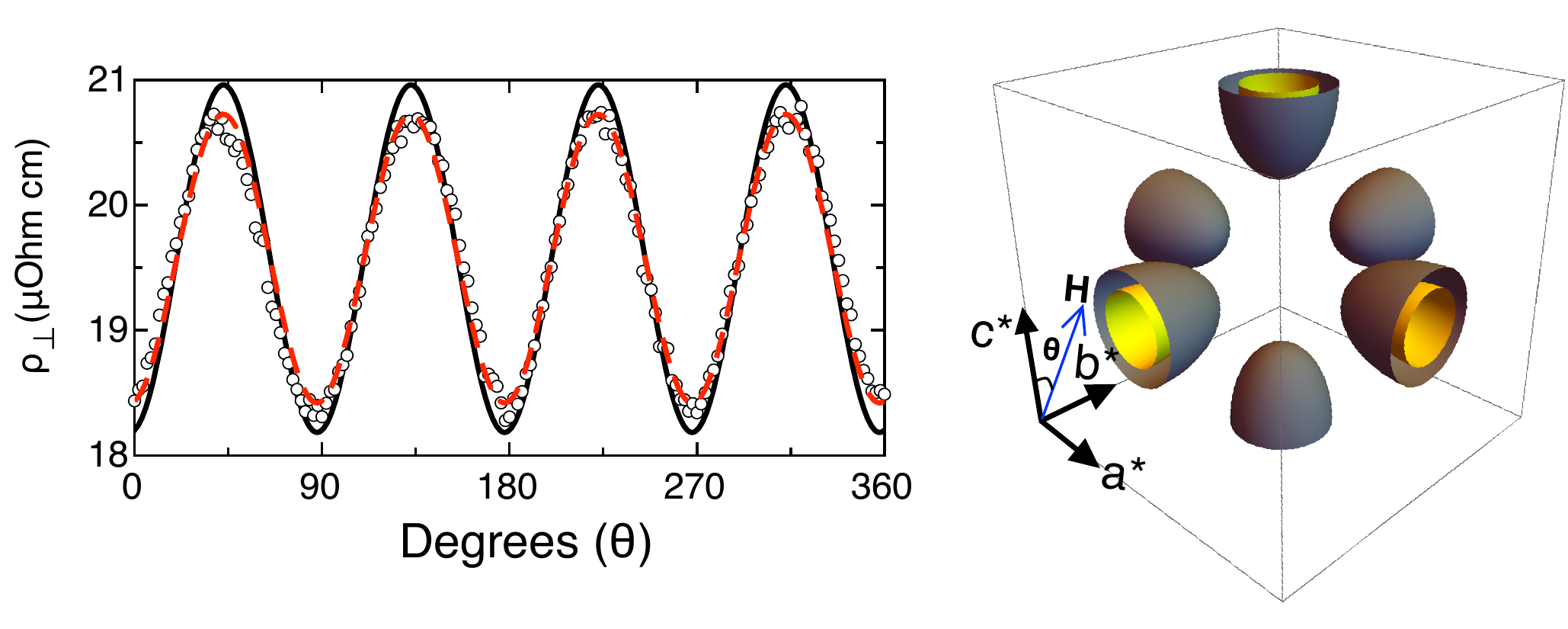}
\caption{ (Left) The angular dependence of the resistivity tensor in a magnetic field of 1~T as calculated from Eq.~\ref{eq:Chambers} is illustrated by the black line. The data are taken from sample \#1 at 4~K and 1~T and fitted with eq. \ref{eq:rho_H_T_theta} (red dash line). (Right) The Fermi surface of \eu, as observed by dHvA~\cite{Goodrich_FS_1998}. The gray outer ellipsoids show the electron pockets and the yellow inner ellipsoids the holes. There are a total of six   half ellipsoids in one unit cell for both electrons and holes.}
\label{fig:Fermi_surface}
\end{figure*}

\andrea{We start the  discussion of our results by presenting the AMRO measurements on the sample with an almost square cross-section (sample \#1 in Table~\ref{tab:demag}) in Fig.~\ref{fig:RvsTheta_cubic}, where the demagnetization effects are smallest. The geometry used in the measurement is schematically shown in Fig.~\ref{fig:RvsTheta_cubic}a). In our convention,  we define the angle $\theta=0$ when the field is applied along $[001]$. The field rotates in the $(100)$-plane  around the direction of the current,  which flows parallel to $[100]$. In order to eliminate any contribution from the Hall effect to the AMRO, we measured the AMRO for both $+H$, and $-H$, and averaged the values.}

\andrea{As can be seen in Fig.~\ref{fig:RvsTheta_cubic}c, for this sample the AMRO signal disappears for temperatures above 50~K in an applied field of 0.5~T. Below 50~K, the strongest AMRO component is a fourfold oscillation $a_4$ from the Fermi surface. This contribution is  expected from the band structure of \eu\, which has a cubic lattice. Such a fourfold oscillation has previously been reported~\cite{urbano_magnetic_2004,Glushkov_2009}. However, for temperatures close to $T_M$, where the specific heat shows a first anomaly~\cite{degiorgi_low-temperature_1997, sullow_structure_1998,sullow_metallization_2000}, and magnetic fields below 1~T, there are two different two-fold contributions the AMRO, which dominate the AMRO response: a smaller two-fold component $a_\mathrm{dem}$ which is commensurate with the fourfold component $a_4$ from the Fermi surface and an $a_2$-component, which is shifted with respect to the $a_4$, and the $a_\mathrm{dem}$-signal. This is particularly visible in the traces taken at 11~K, which were taken in different applied magnetic fields, as shown in Fig.~\ref{fig:RvsTheta_cubic}. Applied fields above 2~T almost completely suppress the two-fold components, as can be seen in Fig.~\ref{fig:RvsTheta_cubic}e) for a temperature of 14~K.}

\andrea{
Thus, a full description of the AMRO requires three components:}
\begin{align}
    \rho(H, T, \theta) =\rho_0\qty(H,T)\Big( 1+a_2 \sin2\theta + a_4 \cos4\theta + a_{\mathrm{dem}}\cos2\theta\Big),
    \label{eq:rho_H_T_theta}
\end{align}
\andrea{where $a_\mathrm{dem}$ is due to demagnetization effects when the sample cross-section deviates from being square, $a_4$ is a four-fold component, and $a_2$ a symmetry breaking contribution. 
This shift result in peak position for the two-fold components at roughly 37$^\circ$. In Eq.~\ref{eq:rho_H_T_theta}, all the values are given as a fraction of the mean resistivity value $\rho_0(T,H)$.}

\andrea{In order to find the correct mechanical zero of the rotator, we first measured the AMRO at 4~K and in a field of 2~T shown in Fig.~\ref{fig:RvsTheta_cubic}d), where the two-fold oscillations are strongly suppressed. We then fitted this  set of data to}
\begin{equation}
    \rho(H, T, \theta) =\rho_0\qty(H,T)\bigg(1+a_4 \cos\big(4(\theta+\gamma)\big)\bigg),
    \label{eq:rho_H_T_zero}
\end{equation}
\andrea{which allowed us to determine the mechanical offset $\gamma$ of $2.3\pm 0.3^\circ$. In the following, we then shifted the origin to this value. A fit of Eq.~\ref{eq:rho_H_T_theta} together with the resulting is shown in Fig.~\ref{fig:RvsTheta_cubic}f) for 14~K in a field of 1~T for sample \#1.} \andrea{We then proceeded to extract the amplitudes for the three components to the AMRO by fitting Eq.~\ref{eq:rho_H_T_theta} for sample \#1 in Table~\ref{tab:demag}. The results are shown in Fig.~\ref{fig:amplitudes_vs_T}. The signal from the Fermi surface $a_4$ is strong below 20~K and increases with increasing field and decreasing temperature, while the amplitude of the symmetry breaking contribution $a_2$ also increases below 20~K. However, is strongest  roughly at $T_\mathrm{C}$ of 12.6~K, and then becomes weaker towards the lowest measured temperatures. Also, unlike $a_4$, $a_2$ goes through a maximum with increasing field, and then vanishes if the field is further increased. A detailed discussion of the demagnetization is given in the Appendix.}

\begin{figure*}
\centering
\includegraphics[width=0.9\textwidth]{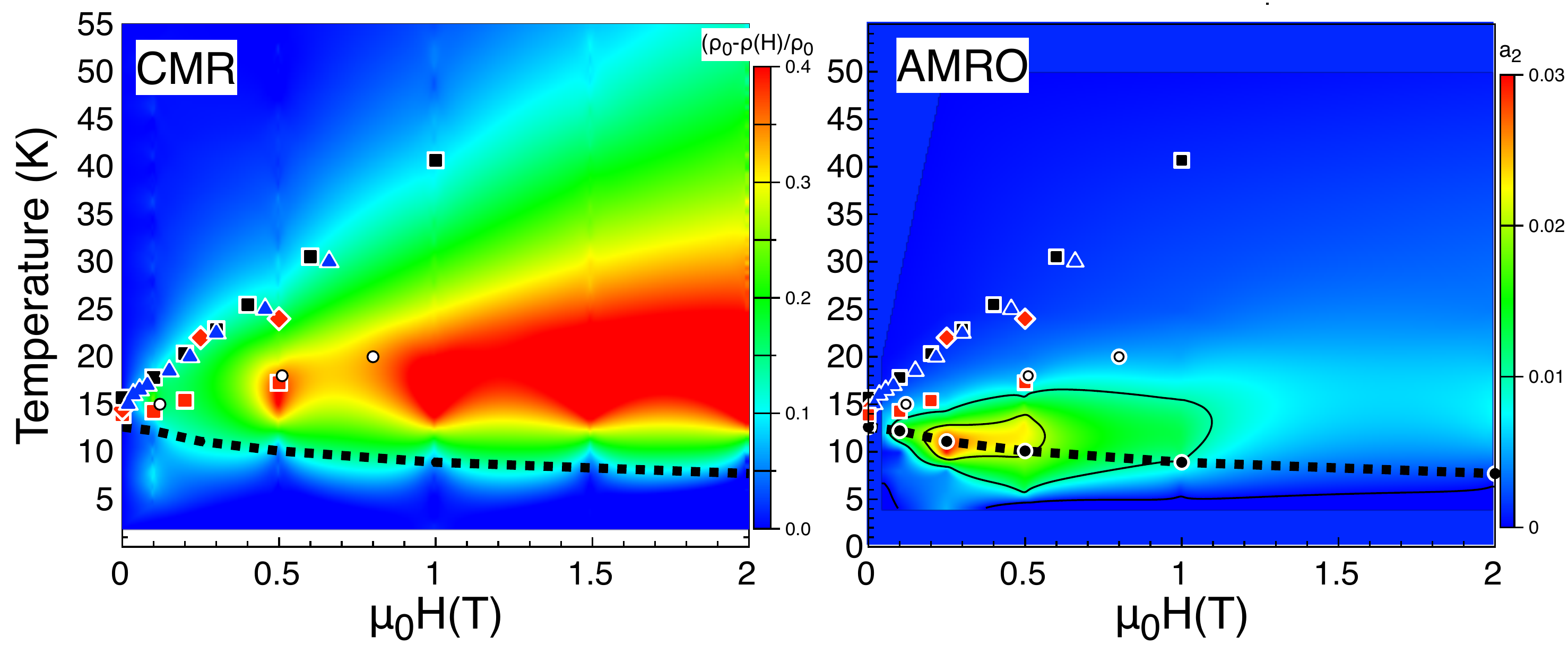}
\caption{\andrea{\textbf{CMR and AMRO phase diagram:} (Full black squares) Hall effect measurements~\cite{zhang_nonlinear_2009}, (Red diamonds) small-angle neutron scattering measurements~\cite{beaudin_2019}, (teal triangles) magnetoresistance peak~\cite{das_magnetically_2012}, (Open black circles) magnetostriction~\cite{manna_lattice_2014}, (Red squares) $T_M$ taken from $\chi"$ measurements (Fig.~\ref{fig:chi}). (Black circles) $T_\mathrm{C}$ from heat capacity found using a fit to mean field theory (Appendix Fig.~\ref{fig:cp}). Left: CMR phase diagram with $(\rho_0-\rho_{001}(H))/\rho_0$ versus temperature, and magnetic fields, where $\rho_0$ is the electrical resistivity in zero field. Right: AMRO phase diagram with the amplitude of the two contributions versus temperature and field. Magnetic polarons are found in the region below the line which is delineated by Hall effect and SANS measurements, and above the
ferromagnetic transition line.}}
\label{fig:phase_diagram}
\end{figure*}

\subsection{Fermi surface contribution $a_4$ to the AMRO}

\andrea{More interesting are the $a_4$ and $a_2$ components to the AMRO. First we are going to  discuss the $a_4$ component, which is due to the fact that \eu\ has an anisotropic Fermi surface, as shown in Fig.~\ref{fig:Fermi_surface}. This finding is in agreement with the results of a previous AMRO study on \eu~\cite{urbano_magnetic_2004}, where the same rotation axis was used. Conventional metals  possess a three-dimensional Fermi surface, which is the position  of the long-lived electronic excitations in reciprocal space  which  determine their electronic properties at low temperatures \cite{Hussey_highTc_2003}. These excitations, as well as the geometry of Fermi surface can be probed by measuring the AMRO.
The AMRO oscillations can  be calculated using the Shockley-Chambers tube integral form of the Boltzmann transport equation, where we have assumed an isotropic mean-free-path \cite{abdel-jawad_anisotropic_2006}.}

The Fermi surface of \eu\ was previously observed by de Haas-van Alphen (dHvA) measurements at 0.4~K~\cite{Goodrich_FS_1998}. The experimentally observed Fermi surface has a cubic symmetry with symmetric ellipsoids at the $X$ points of the Brillouin zone  (see Fig.~3d). A similar Fermi surface was also seen in angle resolved photoemission (ARPES)~\cite{denlinger_bulk_2002}, which agrees well with band structure calculations~\cite{massidda_electronic_1996}. \textcolor{black}{As the dHvA measurements did not indicate a splitting of the Fermi surface, and tunneling experiments indicate a spin-spitting of the Fermi surface only in the ferromagnetic state, we did not take spin-splitting into consideration in our modeling.}

The Fermi surface of \eu\ has an electron pocket radius ratio of 1.8, and a ratio of 1.6 for the holes. The ARPES data~\cite{denlinger_bulk_2002} allows us to estimate the radius of 0.1~\AA$^{-1}$ for the ellipsoids. The temperature dependence of the amplitudes of the dHvA oscillations~\cite{aronson_fermi_1999} gives an effective mass of (0.225$\pm 0.01)m_e$ for the electrons and (0.313$\pm 0.02)m_e$  for the holes. Here, $m_e$ is the electron mass. Using all these parameters, we can calculate the conductivity tensor when a field is applied to the system through the Chambers formula~\cite{chambers_kinetic_1952, Snow_scatteringrate_2001}:
  \begin{align}
  \sigma_{ij}=&\sum_{\alpha}\frac{e^2}{4\pi^3}\frac{m^*_{\alpha}}{\hbar^2 k_{F,\alpha}}\int_{S_{\alpha}} v_{i, \alpha}(\mathbf{k},0)~d^2\mathbf{k} \nonumber \\
  & \times  \int_0^{\infty} v_{j,\alpha}(\mathbf{k},t)e^{-t/\tau}dt \label{eq:Chambers}
  \end{align}
where $S_{\alpha}$ is the Fermi surface sheet associated with band $\alpha$, $k_{F,\alpha}$ is the Fermi momentum of band $\alpha$ defined as $k_{F,\alpha}=\sqrt[3]{3\pi^2 n_\alpha}$, where $n_\alpha$ is the corresponding charge carrier density. The sum is over all occupied bands $\alpha$, where $v_{i, \alpha}$ is the velocity component of band $\alpha$ (either electron or hole),
and $\tau$ is the quasiparticle lifetime.

\textcolor{black}{In \eu, there is no anisotropy in the quasiparticle lifetime, unlike what is observed in some cuprates  \cite{ramshaw2017highTc}. Also, for \eu, the product of cyclotron frequency $\omega_c$  and $\tau$ is greater than one, therefore the quasiparticles on the Fermi surface complete at least one orbit before they scatter~\cite{Hussey_highTc_2003}. This removes some of the complexity in fitting which affects the shape of the AMRO oscillations. Furthermore, since the Fermi pockets in EuB$_6$ are far for the Brillouin zone boundaries, we can neglect large variation in the density of states due to a van Hove singularity~\cite{Matt_ARPES_2015}.}

\andrea{
We performed a theoretical modeling to gain further insight into our AMRO observations by using the relaxation time approximation for the conductivity. AMRO has been instrumental in the early understanding of the band structure of metals~\cite{Kapitza1929,Jones1934}. More recently for example, it provided clear evidence for a nematic state and a symmetry-breaking of the Fermi surface in a high-$T_c$ superconductor~\cite{ramshaw2017highTc}. From Eq.~\ref{eq:Chambers}, a fourfold oscillation in $\rho_{\perp}$ is expected when the magnetic field is rotated in the $b-c$ plane (see Fig.~\ref{fig:Fermi_surface}) due to the $C_4$ symmetry of the Fermi surface. The calculation at a applied field of 1~T gives an amplitude $a_4$ of -0.07 which is lower than the amplitude of -0.06 found at 4~K. This is valid since the amplitude keep decreasing with temperature (see Fig.~\ref{fig:amplitudes_vs_T}), and the temperature in the calculations was zero. The resulting 4-fold oscillatory component is shown in Fig.~\ref{fig:Fermi_surface} 
The overall constant resistivity differs strongly with the experimental data, which is due mainly by approximation and initial parameters. The value was scaled in Fig. \ref{fig:Fermi_surface} to make the comparison. Further improvement to the numerical calculation will be done to investigate the symmetry breaking.
}

\begin{figure}
\includegraphics[width=0.8\linewidth]{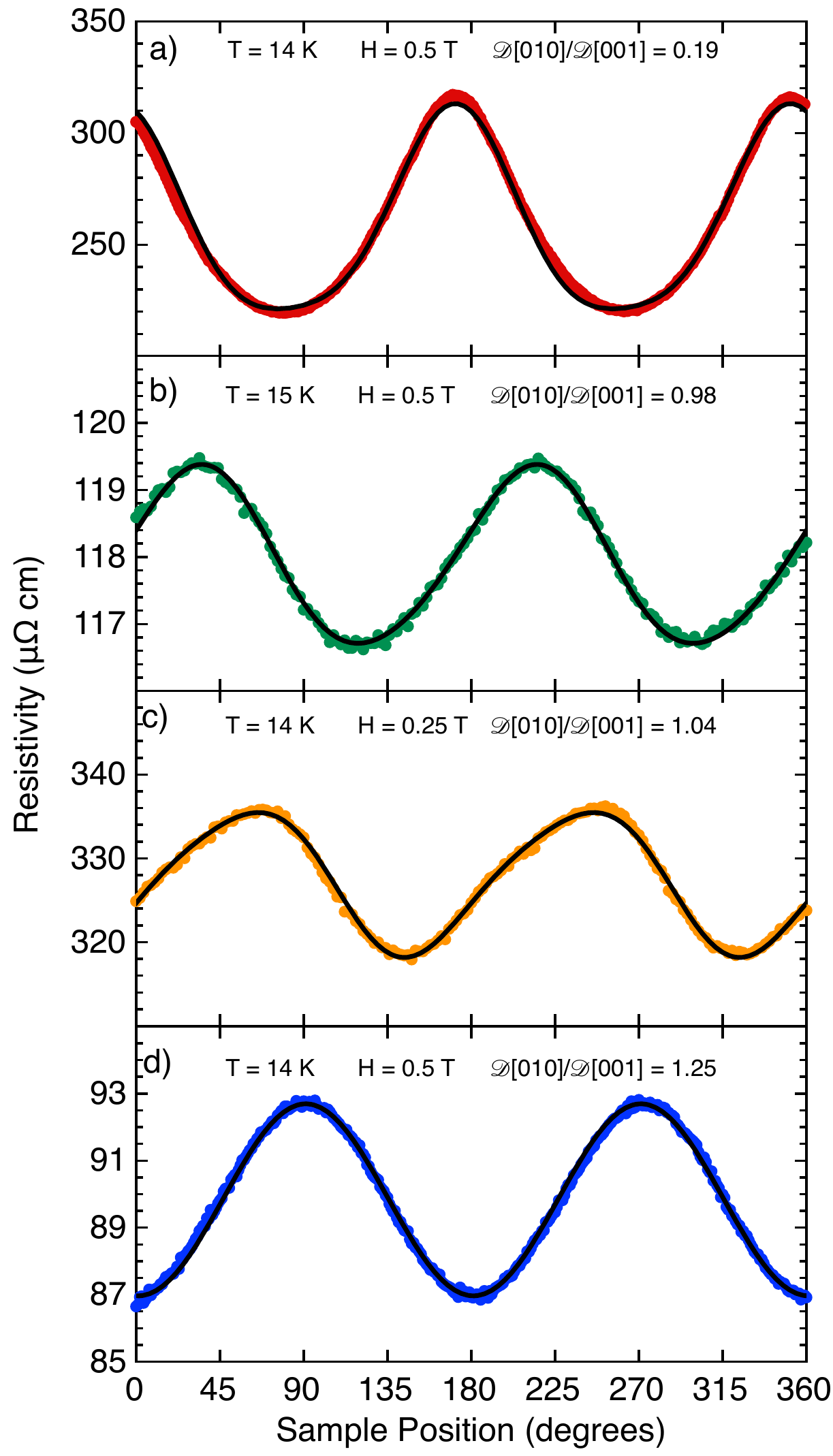}
\caption{ Comparison of the AMRO signal around T$_\mathrm{M}$ and at  applied fields of 0.25, and 0.5~T for a) sample \#6 b) sample\# 1, c) sample \# 2 and d) sample \# 4. The black lines are a fit to Eq.~\ref{eq:rho_H_T_theta}.}
\label{fig:signal_comparaison}
\end{figure}

\subsection{Symmetry breaking contribution $a_2$ to the AMRO}
\andrea{Finally, we will discuss the new symmetry breaking contribution $a_2$. The symmetry breaking two-fold contribution is not due to aluminum inclusion from the flux growth, as those would contribute a fourfold pattern, which is in phase with the $a_4$ component of the AMRO as discussed in further detail in App.~\ref{app:alu}. The $a_2$ contribution  breaks the inversion symmetry, which is in agreement with the spin-split Fermi surface observed in a recent ARPES study by \cite{gao_time-reversal_2021}. The tendency of divalent hexaborides towards ferromagnetism associated with chirality, which breaks the cubic symmetry, was also discussed in the context of the formation of an excitonic insulator in La doped CaB$_6$~\cite{zhitomirsky_electron-hole_2000,barzykin_ferromagnetism_2000-1,balents_ferromagnetism_2000}.}

\andrea{The colors in Fig.~\ref{fig:phase_diagram} show the strength of the symmetry breaking component $a_2$ of the AMRO, as determined from the size of the two-fold contribution to $\rho_\perp$ varying as \andrea{$\sin(2\theta)$, where $\theta$ is the angle of the applied field in the plane perpendicular to the current (Fig~\ref{fig:RvsTheta_cubic}a))}. The symmetry breaking component appears below the $T_M$ line in the phase diagram demarked by magnetostriction~\cite{manna_lattice_2014} and $\chi"$ given in Sec.~\ref{sec:chi}. Also, this signal is only seen in nearly cubic samples. Fig.~\ref{fig:signal_comparaison} shows how non-cubic samples deviate from the $\sin2\theta$ because of demagnetization. This effect is explained in more detail in App.~\ref{app:demag}.}

\subsection{Magnetic susceptibility}
\label{sec:chi}

In order to better map out the $(H, T)$ phase diagram, and delineate the region where we observe a quantum nematic in the AMRO, we carried out the AC susceptibility. 
The AC susceptibility is a complex value and reads
$\chi_\mathrm{AC} = \chi' - i\chi"$, where the real component $\chi'$, related to the reversible magnetization process, and the imaginary component $\chi"$ is related to losses due to the irreversible magnetization process and energy absorbed from the field~\cite{balanda_ac_2013}. One such example are   the vortices in type-II superconductors, which are topological defects of an homogeneous order parameter. 

\begin{figure}
\includegraphics[width=0.9\linewidth]{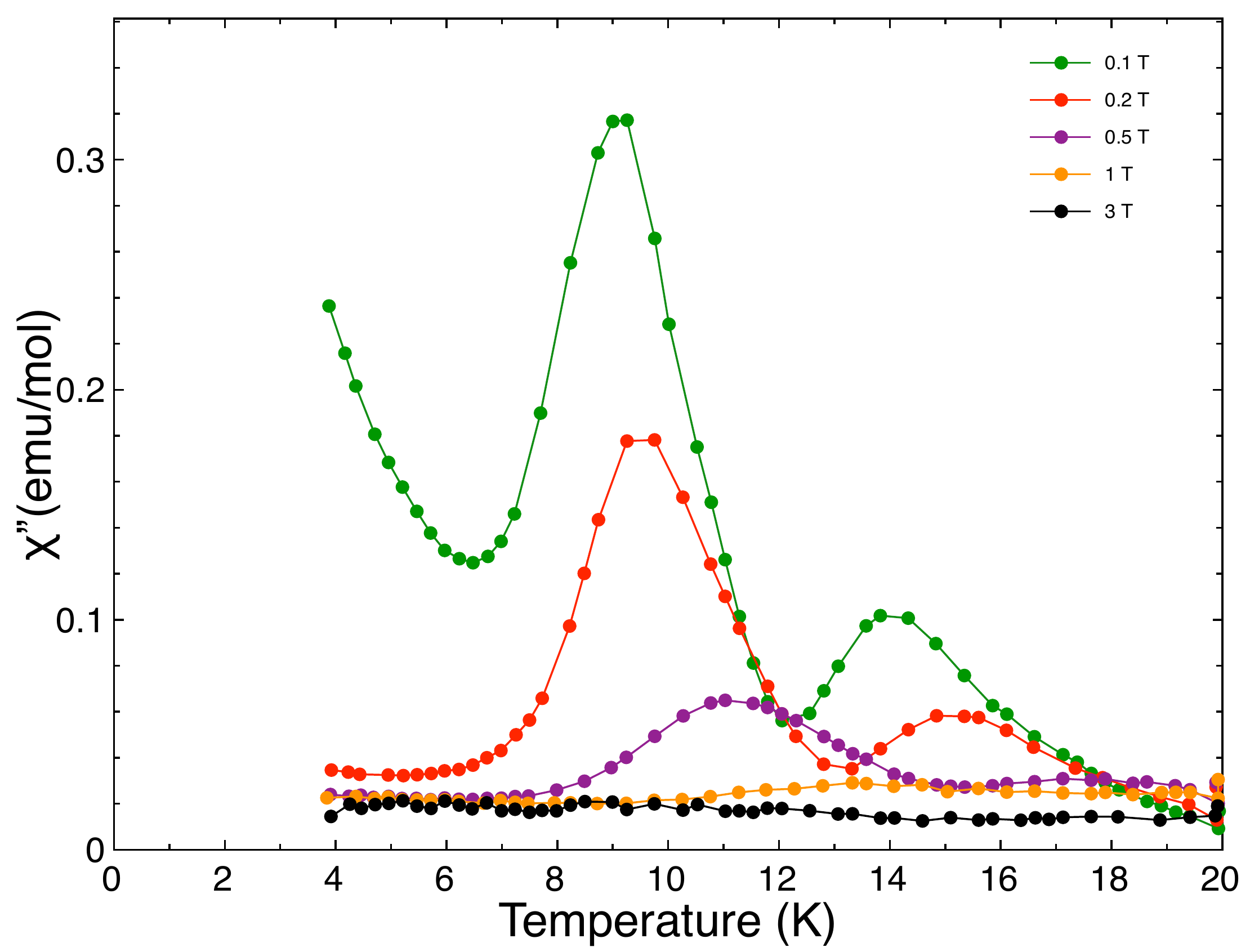}
\caption{\textbf{Imaginary part of the magnetic susceptibility} $\chi''$ versus temperature for different applied fields. The second maximum in $\chi''$ coincides with where the AMRO is strongest.}
\label{fig:chi}
\end{figure}

Fig.~\ref{fig:chi} shows the second order magnetic susceptibility $\chi''$. The peak in $\chi''$ at the lower temperature corresponds to the ferromagnetic transition, while  the  peak at higher temperatures corresponds to where the signal of magnetic polarons is strongest. The positions of the high temperature  peaks are shown as red squares in Fig.~\ref{fig:phase_diagram}. This phase boundary  coincides with the increase in anisotropic resistivity signal which suggests a strong connection between nematicity and magnetic polarons in \eu.

\section{Outlook}
The role electronic phase separation and magnetic polarons~\cite{kivelson_electronic_1998,2006stripes} play in high temperature superconductivity clearly motivates the need for a model system for studying magnetic polarons. In \eu, the regions of the $(H, T)$ phase diagram where SANS indicates magnetic polarons and where AMRO displays quantum nematicity coincide, providing strong evidence that both nematicity and magnetic polarons originate from the same electronic correlations. This makes our results on \eu\ particularly important, as unlike the high temperature superconductors, \eu\ possess a high degree of structural order without a nearby lattice instability~\cite{booth_local_2001}. \eu\ is thus an excellent model system to study quantum nematicity. Further, the presence of both an electronic nematic and colossal magnetoresistance suggests that \eu\ can be used as a novel platform for spintronic devices~\cite{awschalom_challenges_2007}. This promises a way to harness strong electronic correlations for spintronic applications, and motivates the search for other materials with magnetic polarons.


\begin{acknowledgements}
The research at Universit\'e de Montr\'eal received support from the Natural Sciences and Engineering Research Council of Canada (Canada). W.\,W.-K. was in addition supported by the Fondation Courtois, a Canada Research Chair, and a ''\'Etablissement de nouveaux chercheurs et de nouvelles chercheuses universitaires'' grant from the FRQNT. The work at the MPI-CPFS was enabled through a DAAD grant. \andrea{The authors would like to thank Prof. S. S\"ullow from the TU Braunschweig in Germany for sharing his magnetoresistance data on \eu\ with us.}
\end{acknowledgements}

\appendix
\section{Demagnetization contribution to the AMRO}

\begin{figure*}
    \centering
    \includegraphics[width=\textwidth]{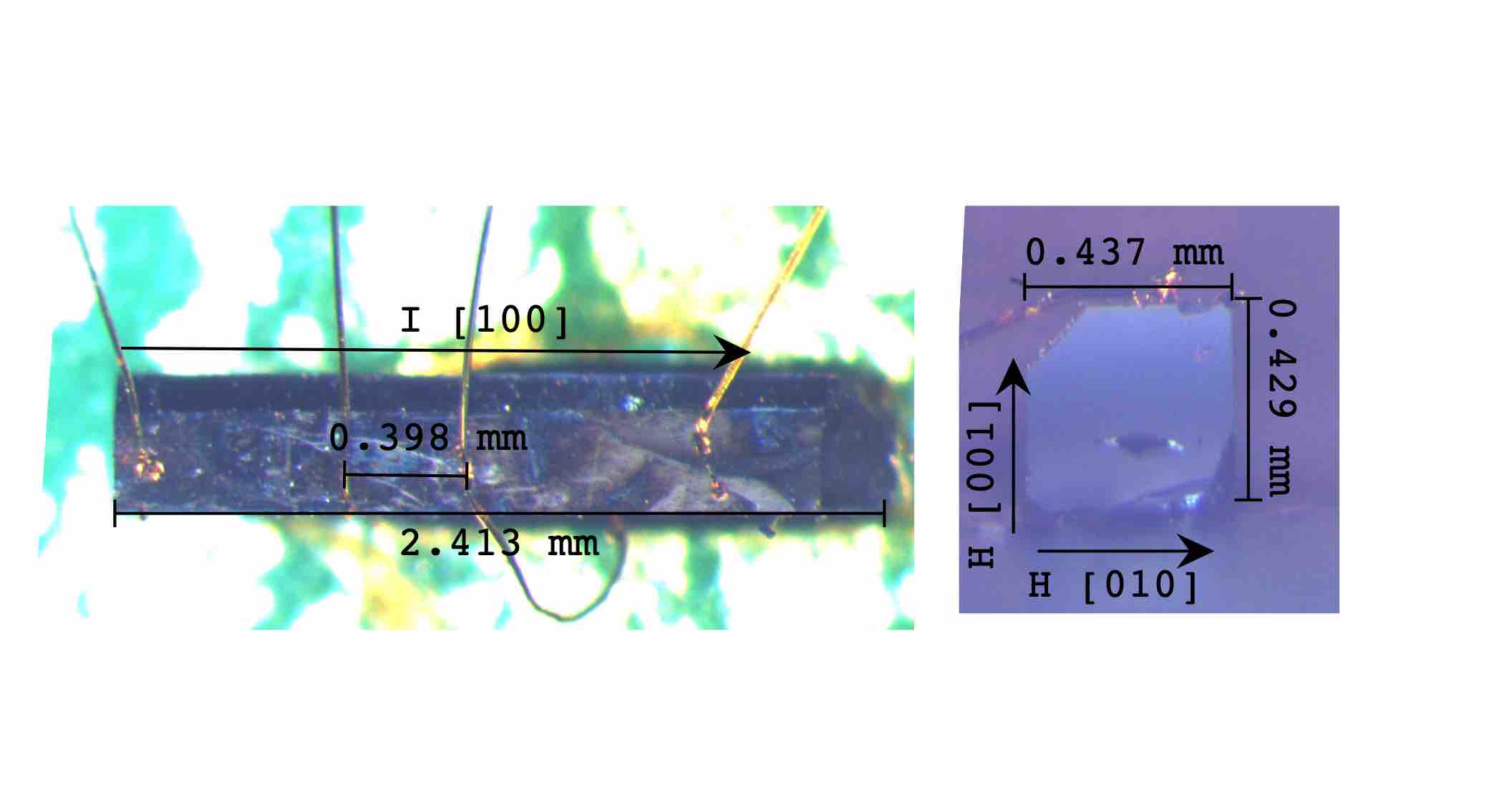}
    \caption{\textbf{Dimensions of   sample \#1:} For this sample, the demagnetization factors  are $\mathcal{D}_{100}=0.080$, $D_{010}=0.46$, and $\mathcal{D}_{001}=0.46$. For the contacts we spot-welded gold wires with a diameter of 25~$\mu$m to the sample.}
    \label{fig:sample}
\end{figure*}

\label{app:demag}
\begin{figure}
\includegraphics[width=0.9 \linewidth]{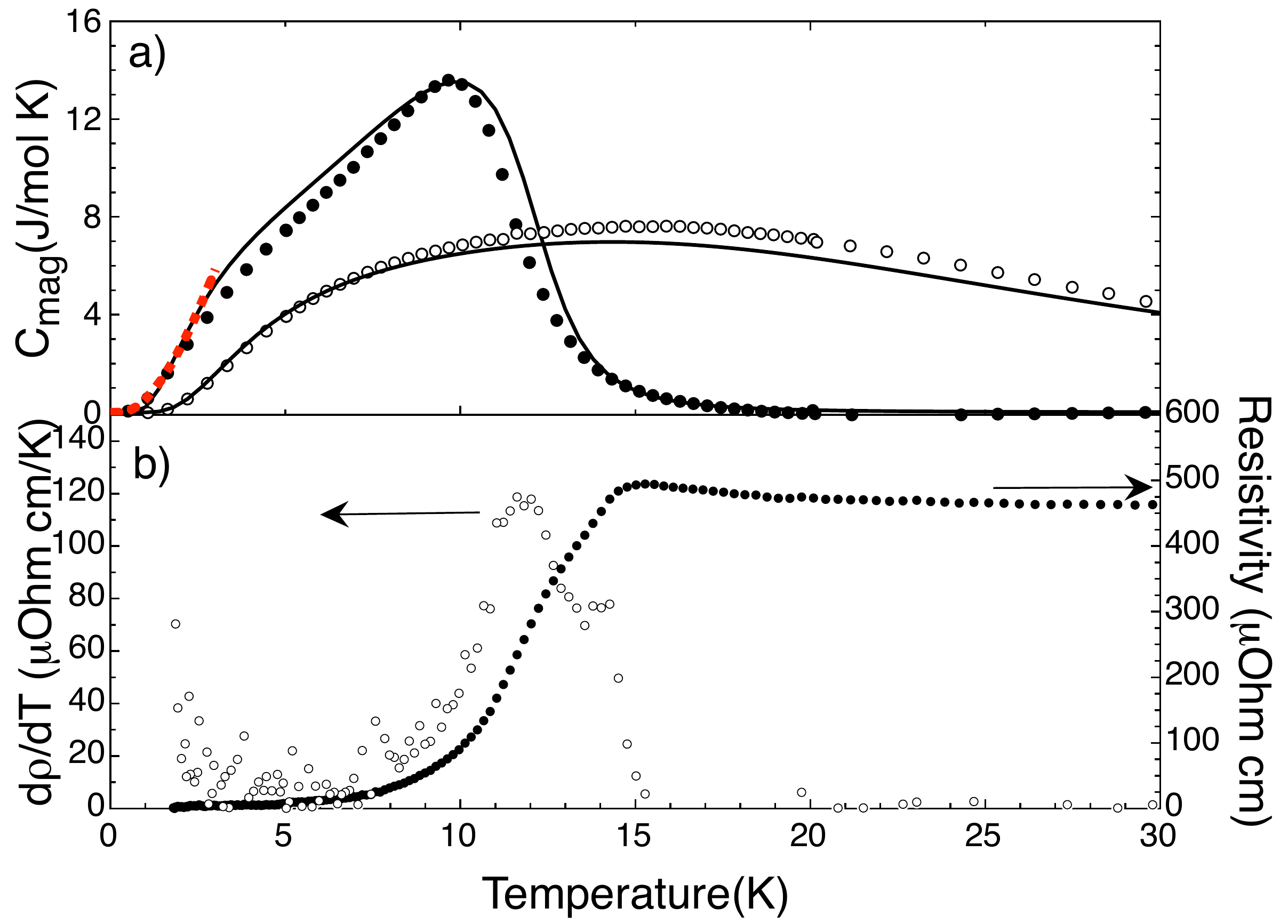}
\caption{\textbf{Specific heat and electrical resistivity of EuB$_6$} \textbf{a} Magnetic specific heat capacity $C_M=C_p-C_{\mathrm{ph}}$ after subtraction of the phonon contribution $C_{\mathrm{ph}}$ of EuB$_6$ at zero field. The curves correspond to   0~T (full circles) and  5~T (empty circles). The black line is a fit to the mean field  model for the heat capacity~\cite{rodriguez_temperature_2005-1}. A field of 0.1~T was used for the zero field data. The red dashed line is a fit to a spin wave contribution, which is proportional to $T^{3/2}e^{-\Delta/T}$\cite{coqblin_electronic_1977}. It gives a gap of $\Delta=1.11\pm0.03$~K similar to the one reported by NMR techniques\cite{ambrosini_nmr_1999}. Contrary to previous work~  \cite{degiorgi_low-temperature_1997}, we only see a shoulder at the onset of $T_{\mathrm{M}}$ instead of a second peak in $C_M$. \textbf{b} Electrical resistivity $\rho$ at zero field, and its temperature derivative $\partial \rho/\partial T$ versus temperature for the sample used in the main text.}
\label{fig:cp}
\end{figure}

\andrea{The demagnetization contribution can be calculated, if we know the magnetoresistance as a function of the magnetization (see for example~\cite{sullow_metallization_2000}), using the formula for the demagnetization factor~\cite{aharoni_demagnetizing_1998}, we can then calculate this contribution. 
The contribution $a_\mathrm{dem}$ from the demagnetization to the  AMRO depends on both the internal magnetic field $B$ and the magnetoresistance $\pdv{\rho}{B}$. The demagnetizing factor $\mathcal{D}$ for rectangular ferromagnetic prisms was given by~\cite{aharoni_demagnetizing_1998}. If for example the sample does not have a square cross-section (for example, $t([001]) < w([010])$, as shown in Fig.~\ref{fig:RvsTheta_cubic}a).), then if $\mathcal{D}_{001}>\mathcal{D}_{010}$, rotating the sample around the direction of the current flow ($[100]$) from the $[001]$ direction to the $[010]$ direction is equivalent to an increase in the magnetic field ($\Delta B = B_{001}-B_{010} < 0$). As the magnetoresistance $\pdv{\rho}{B}$ is negative, the effect of demagnetization would be that $\Delta \rho_{\mathrm{demag}}\approx \pdv{\rho}{B} \Delta B > 0$. Thus for such a sample, the $\rho(B\parallel[001])>\rho(B\parallel[010])$.}
\andrea{
For a given direction, the internal field in a sample is defined as~\cite{coey_magnetism_2001}:}
\begin{align}
B / \mu_{0}&=H_0+M-H_{\mathcal{D}}\\
                    &=H_{0}+M-\mathcal{D}M\\
                    &=H_0+\qty(1-\mathcal{D}) M\qty(T,H,H_{\mathcal{D}})\qquad,
\label{eq:heff}
\end{align}
\andrea{Here, $H_0$ is the applied field and $M$ is the magnetization as a function of $H$ and $H_{\mathcal{D}}$.
The difference in the internal magnetic field, for when the magnetic field is applied along a different geometric axis, can then be calculated  using demagnetization factor  \cite{aharoni_demagnetizing_1998}. If we know the magnetization for a given applied field, we can then compute the internal field, which is the sum of the applied magnetic field plus the induced magnetization corrected by the demagnetization. We can find the induced magnetization for a given  applied field with the help of Eq.~\ref{eq:heff}, using a mean field approximation~\cite{rodriguez_temperature_2005-1}. The use of a mean field approximation for finding the magnetization is justified, since it describes the field dependence of the specific heat well, as  is shown in Fig.~\ref{fig:cp}.}

\andrea{To find the magnetization, we numerically solve the following set of equations \cite{rodriguez_temperature_2005-1}:} 
\begin{align}
    M &= M_0B_J(x)\\
    B_J(x)&= \frac{2 J + 1}{2 J} \coth\qty(\frac{2 J + 1}{2 J} x) - \frac{1}{2J}\coth\qty(\frac{x}{2J})\\
   x &= \frac{gJ\mu_{\mathrm{B}}}{k_{\mathrm{B}}T}\bigg(\mu_0 H_0 + (\lambda-\mathcal{D})\mu_0 M\bigg)\\
   \lambda &= \frac{3 k_{\mathrm{B}}T_{\mathrm{C}}}{Ng^2\mu^2_{\mathrm{B}}\mu_0 J(J+1)}
\end{align}
\andrea{Here, $B_J\qty(x)$ is the Brillouin function, $J$ is the total angular momentum   of $\frac{7}{2}$ of \eu, $T_\mathrm{C}$ the Curie-Weiss temperature of 12.6~K, the $g$-factor is 2 and $N$ is the number of magnetic atoms per volume. However, in order to be able to calculate the demagnetization contribution $a_\mathrm{dem}$, the magnetization dependence of the electrical resistivity has to be measured.}

\begin{figure}
    \centering
    \includegraphics[width=0.9\linewidth]{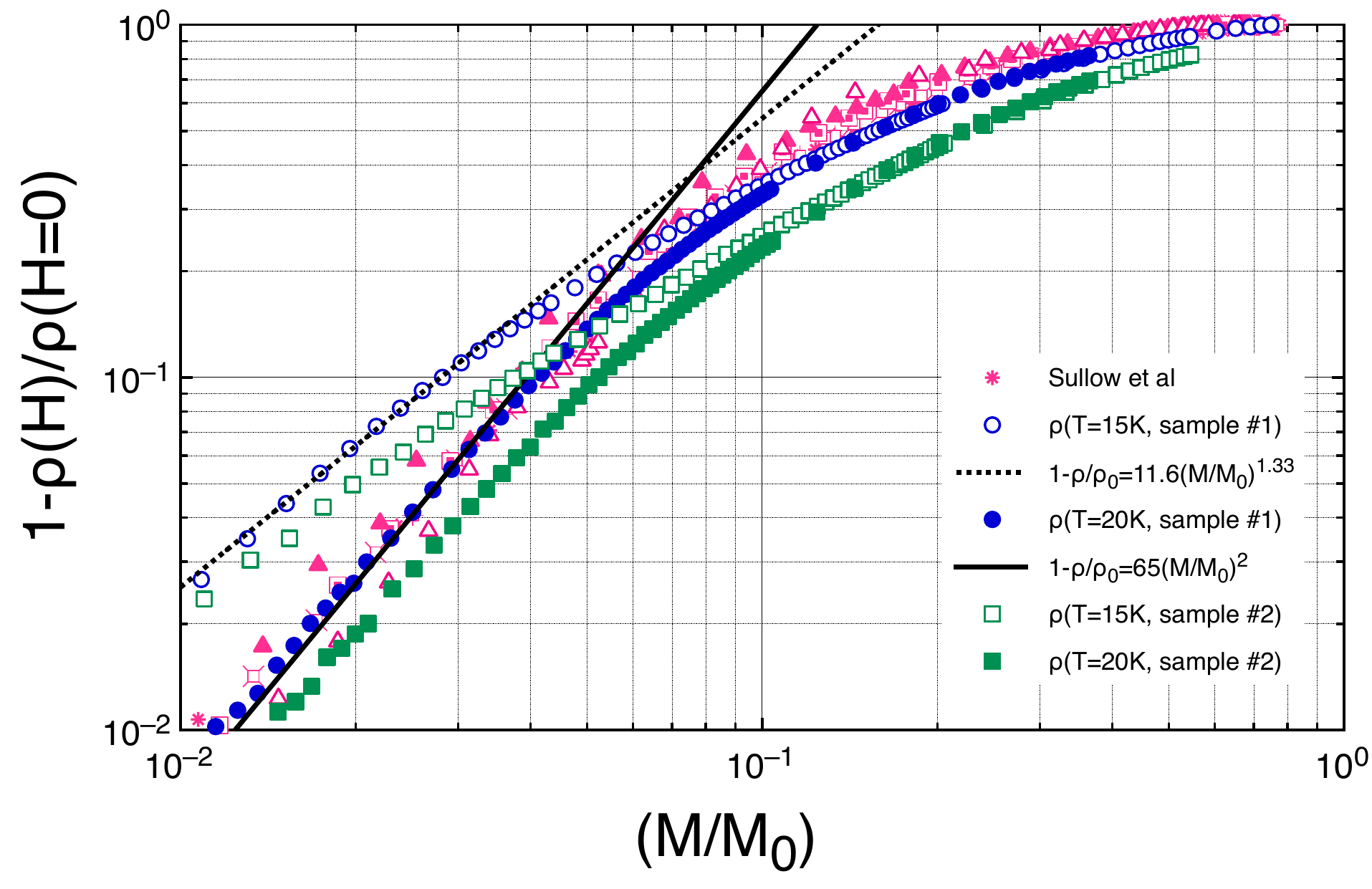}
    \caption{\andrea{\textbf{Magnetoresistance versus magnetization} Here, $M_0$ is the value for the saturation magnetization of \eu\, and $\rho\qty(H)$ the electrical resistivity in an applied field  of $H$. The blue circles show the magnetoresistance of the sample with a square cross-section (sample \#1)  as $1-\rho(H)/\rho(H=0)$ at 15~K(open), and at 20~K (closed). The green squares show the magnetoresistance of the sample with a square cross-section (sample \#2)  as $1-\rho(H)/\rho(H=0)$ at 15~K(open), and at 20~K (closed). The calculation of $M$ accounts for demagnetization. For all, the magnetic field was applied along  $[001]$.  The solid, dotted and dash lines are a fit to equations a power law. The solid pink points are data reported previously by S\"ullow \emph{et al.}~\cite{sullow_magnetotransport_2000}, which follows  $1-\rho(H)/\rho(H=0)\sim 75\qty(\frac{M}{M_0})^{2}$.}}
    \label{fig:drdM}
\end{figure}

\begin{figure}
    \centering
    \includegraphics[width=0.9 \linewidth]{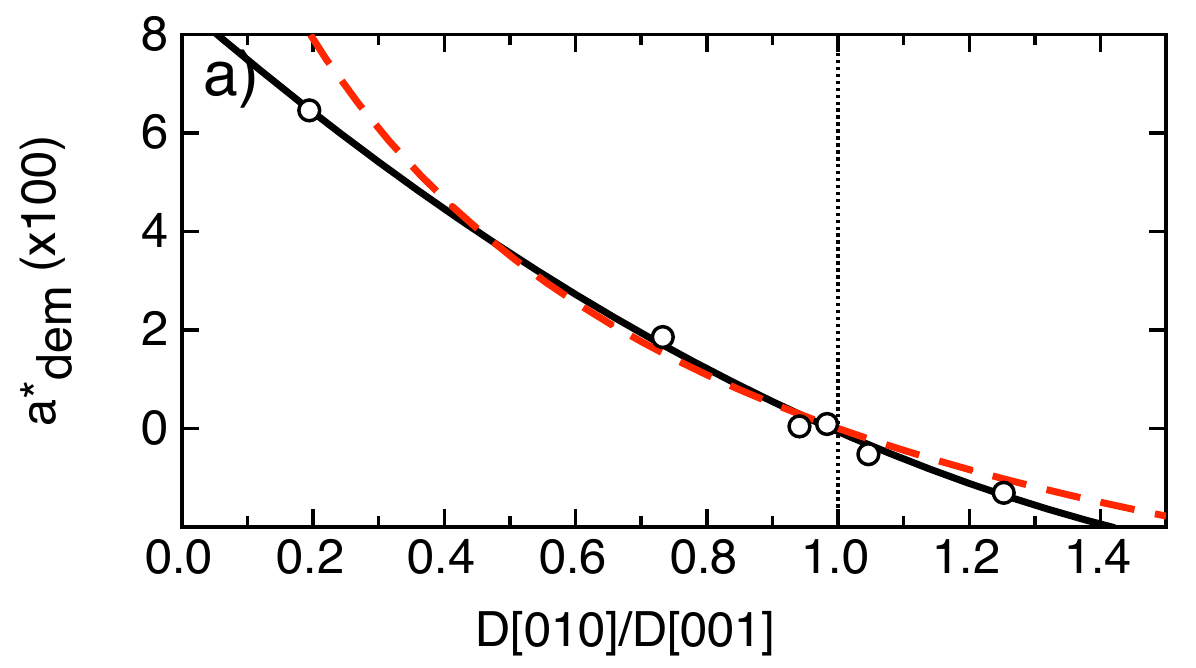}
    \caption{\andrea{\textbf{Two-fold contribution to the AMRO} The open circles show the amplitude $a^*_{dem}$ retrieved from the fit of Eq.~\ref{eq:rho_H_T_theta} versus the ration of the demagnetization factors, $\frac{\mathcal{D}_{010}}{\mathcal{D}_{001}}$ for  all samples at 15~K and in an applied field of 0.5~T. The amplitude was normalized as $a^*_\mathrm{dem}=\frac{\rho_0(T=15~\mathrm{K}, H = 0.5~\mu_0~\mathrm{T})}{\rho_0(T = 15~\mathrm{K}, H=0)} a_\mathrm{dem}$ for a comparison with the data. The solid line a fit of $a_\mathrm{dem} \propto \qty(\frac{\mathcal{D}_{010}}{\mathcal{D}_{001}})^2$. The dashed line is the calculated amplitude using the demagnetization and the CMR as given by Eq.~\ref{eq:rho_of_H} as described in the text.} }
    \label{fig:demag}
\end{figure}

\andrea{Madjumdar and Littlewood proposed that the magnetoresistance in metallic ferromagnets, and doped
magnetic semiconductors should be negative, and proportional to the square field induced magnetization~\cite{majumdar_dependence_1998}. In their theory, this behavior is due to the suppression of the magnetic fluctuations by the applied magnetic field. Such a behavior was reported by~\cite{sullow_metallization_2000}, i.e., $ 1-\frac{\rho(H)}{\rho(H=0)}\sim\qty(\frac{M}{M_0})^{2}$. We carried out   magnetoresistance measurements on an \eu\ sample with an almost square cross-section (sample \#1 in Table~\ref{tab:demag}, with $\mathcal{D}_{010}/ \mathcal{D}_{001} = 0.983$, as shown in Fig.~1 of the Supplementary Materials). The results of these measurements are shown in Fig.~\ref{fig:drdM} for 15~K and 20~K. For comparison we also show data extracted  from~\cite{sullow_magnetotransport_2000}. Our data does follow a power-law:}
\begin{align}
    1-\frac{\rho(H)}{\rho(H=0)} &= 65\qty(\frac{M(H)}{M_0})^{2} & \mathrm{for} & & T&=20~\mathrm{K}\qquad\\
    1-\frac{\rho(H)}{\rho(H=0)} &= 11.6\qty(\frac{M(H)}{M_0})^{1.33} & \mathrm{for} & & T&=15~\mathrm{K}\quad,
    \label{eq:rho_of_H}
\end{align}
\andrea{and agrees with the exponent of two at low field above $T_{\mathrm{M}}$. Using the factor found in equation 11 and the charge carrier concentration, we can determine the position of our measurements in the plot presented by Madjumdar and Littlewood. Our factor $C$ is equal to 65, our charge carrier concentration is 5$\times10^{19}$ cm$^{-3}$ at low field and temperature of 20~K and our magnetic correlation length is 4.18~\AA. The charge carrier concentration was taken from Hall coefficient measurements. These values take us in the same region as other \eu\ experiments~\cite{sullow_magnetotransport_2000}. The first indication that $a_{\mathrm{dem}}$ is indeed due to demagnetization comes from the fact that it is commensurate with the $a_4$ component from the Fermi surface. As this component is linked to the shape of the cross-section, and the crystal habit is given by the cubic symmetry of \eu, this is the expected behavior. We used Eq.~\ref{eq:rho_of_H} to calculate the two-fold AMRO component $a_\mathrm{dem}$ due to a combination of demagnetization effects and a large CMR in \eu. The result of this calculation is shown in Fig.~\ref{fig:demag} as the  red dashed line for a temperature of 15~K, and in an applied magnetic field of 0.5~T. Here, we calculated $a_\mathrm{dem}$ as  function of the ratio of the demagnetization factor $\frac{\mathcal{D}_{010}}{\mathcal{D}_{001}}$ using a sample length $L$ of 1~mm.}

\section{Aluminum inclusions}
\label{app:alu}
Previous interpretations of dHvA Fermi surface measurements of \sm\ were plagued by aluminum inclusions, see Refs.~\cite{phelan_chemistry_2016,thomas_quantum_2019}. It was found that the aluminum inclusions are epitaxial single crystals co-oriented with the $(100)$ direction \sm. We would expect the same growth direction for aluminum inclusions in \eu. Such inclusion, if present, would then lead to four-fold AMRO pattern~\cite{balcombe_magneto-resistance_1963}, with the same angle dependence as observed at high temperatures and high magnetic fields in \eu. Thus,  the lowering of the symmetry in the AMRO from four-fold to two-fold in \eu\ cannot be explained by the presence of epitaxial aluminum inclusions.

\section{AMRO calculations}
In Eq.~\ref{eq:Chambers} of the manuscript we calculated a numerical integral over the Fermi Surface $S$, which is shown in Fig.~3 of the manuscript. The total Fermi energy for one closed pocket (for example the electron band at position \textsc{X}) is~\cite{ciftja_impact_2019}:
\begin{align}
E_F &=\frac{V}{(2 \pi)^{3}} \frac{\hbar^{2}}{2 m^*} \quad \iiint_{\{\frac{k_{x}^{2}}{k_{a}^{2}}+\frac{k_{y}^{2}}{k_{a}^{2}}+\frac{k_{z}^{2}}{k_{b}^{2}} \leq 1 \}} d k_{x} d k_{y} d k_z\nonumber \\
\times &\left(k_{x}^{2}+k_{y}^{2}+k_z^2\right)\\
E_F/N&=\frac{\hbar^2}{2m^*}k_F^2 \qquad \mathrm{, where}\\ k_F^2&=\frac{4}{15}(2k_a^2+k_b^2)\qquad,
\end{align}
where $k_a$ is the minor axis, and $k_b$ is the major axis  of the elliptical pockets. These axes are linked to the volume of the Fermi volume,  and hence, to the charge carrier density. So, we can obtain the minor axis from the ratio of the radii   $R^*=k_b/k_a$, and the density of charge carriers $n$ as:
\begin{equation}
    k_a=\frac{\sqrt[3]{3\pi^2n}}{\sqrt{\frac{4}{15}(2+R^{*2})}}
\end{equation}
Here, $n$ and $R^*$ will be different for the two bands (electrons and holes).
The velocity of the charge carriers was calculated as follows:
\begin{equation}
    \mathbf{v}(\mathbf{k})=\frac{1}{\hbar}\nabla_{\mathbf{k}}E(\mathbf{k})
\end{equation} and
\begin{equation}
    E(\mathbf{k})= \frac{\hbar^2}{2m^*}\mathbf{k}^2
\end{equation} where $\mathbf{k}$ is a wave vector on the Fermi surface. 
For the propagation of the particle along a trajectory on the Fermi surface, we use the following equation of motion:
\begin{equation}
    \hbar \frac{d \mathbf{k}}{\mathrm{d} t}=q \mathbf{v}(\mathbf{k}) \times \mathbf{B}(\theta, \varphi),
\end{equation}
where $\mathbf{B}$ is the applied magnetic field at specific angles $\theta$ and $\varphi$. The integral over time was taken from 0 to 9$\tau$, giving time for the particle to do more than one revolution around the Fermi surface and for the integral to converge. In the case of \eu\ the scattering rate $1/\tau$ reported by Ref.~\cite{Snow_scatteringrate_2001} $1/\tau=8.5$~cm$^{-1}=4.05 \times 10^{10}$~s$^{-1}$ is smaller than the cyclotron frequency ($\omega_c=\frac{e B(1~\mathrm{T})}{2\pi m^*}=1.2\times 10^{11}$~s$^{-1}$). This means that in \eu\ the electrons complete more than one revolution around the Brillouin zone before they scatter. Also, in \eu\ the Fermi pockets are far from the Brillouin zone boundaries. As a consequence, for \eu\ the contribution from an anisotropic scattering rate can be neglected.  This is very different from the situation in Tl$_2$Ba$_2$Cu$_1$O$_{6+\delta}$~\cite{analytis_angle-dependent_2007}. In order to speed up calculations, we also assumed a uniform density of the charge carriers.

\begin{table}
\centering
    \begin{tabular}{c|c|c}
       Values  & Electron band & Hole band \\
       \hline
        $R^*$\cite{aronson_fermi_1999} & 1.6 & 1.8 \\
        $m^*$\cite{wigger_electronic_2004} & 0.24 & 0.29 \\
       $\tau$\cite{Snow_scatteringrate_2001}& 8.5 cm$^{-1}$ & 8.5 cm$^{-1}$ \\
        $n$\cite{wigger_electronic_2004} & 6.7$\times$ 10$^{19}$ cm$^{-3}$  & 6.1$\times$ 10$^{19}$ cm$^{-3}$\\
        $n$\cite{zhang_spin-dependent_2008} & 3.23$\times$ 10$^{19}$ cm$^{-3}$  & 3.05$\times$ 10$^{19}$ cm$^{-3}$\\
        $n_{calc}$ & 5$\times$ 10$^{19}$ cm$^{-3}$  & 5$\times$ 10$^{19}$ cm$^{-3}$\\
    \end{tabular}
    \caption{Experimental data used for the AMRO calculations}
\end{table}



\newpage
\bibliographystyle{apsrev4-2}

\bibliography{EuB6_biblio_amro.bib}

\end{document}